\documentclass[pra,epsfig,floats,twocolumn,superscriptaddress,amssymb,floatfix,showpacs]{revtex4}
\usepackage{graphicx}

\begin{document}
\title{Effect of the Heterogeneity of Metamaterials on Casimir-Lifshitz Interaction}

\author{Arash Azari}
\affiliation{Department of Physics and Astronomy, University of
Sheffield, Sheffield S3 7RH, UK}

\author{MirFaez Miri}
\email{miri@iasbs.ac.ir}
\affiliation{Department of Physics, University of Tehran, P.O. Box
14395-547, Tehran, Iran}

\author{Ramin Golestanian}
\email{r.golestanian@sheffield.ac.uk}
\affiliation{Department of Physics and Astronomy, University of
Sheffield, Sheffield S3 7RH, UK}

\date{\today}

\begin{abstract}
The Casimir-Lifshitz interaction between metamaterials is studied using
a model that takes into account the structural heterogeneity of the dielectric
and magnetic properties of the bodies. A recently developed perturbation
theory for the Casimir-Lifshitz interaction between arbitrary material bodies
is generalized to include non-uniform magnetic permeability profiles, and used
to study the interaction between the magneto-dielectric heterostructures
within the leading order. The metamaterials are modeled as two dimensional
arrays of domains with varying permittivity and permeability. In the case of two
semi-infinite bodies with flat boundaries, the patterned structure of the material
properties is found to cause the normal Casimir-Lifshitz force to develop an oscillatory
behavior when the distance between the two bodies is comparable to the wavelength of the
patterned features in the metamaterials. The non-uniformity also leads to the emergence
of lateral Casimir-Lifshitz forces, which tend to strengthen as the gap size becomes smaller.
Our results suggest that the recent studies on Casimir-Lifshitz forces between metamaterials,
which have been performed with the aim of examining the possibility of
observing the repulsive force, should be revisited to include the effect of the
patterned structure at the wavelength of several hundred nanometers that coincides with
the relevant gap size in the experiments.
\end{abstract}


\maketitle
\section{Introduction}

Despite nearly six decades of research after the original works of Casimir \cite{Casimir48}
and Lifshitz \cite{Lifshitz}, the dependence of Casimir-Lifshitz force between bodies on
their geometrical and material properties is still a subject of ongoing investigation \cite{Klim-etal-RMP09}. The effect of geometry has been studied using a variety of techniques,
which include perturbative expansion around ideal geometries \cite{GK,EHGK,lambrecht} and in
dielectric contrast \cite{barton,ramin,buhmann,rudi,milton}, semiclassical \cite{semiclass} and classical
ray-optics \cite{Jaffe} approximations, multiple scattering \cite{balian,klich} and multipole
expansions \cite{multipole1,multipole2,multipole3}, world-line method \cite{gies}, exact numerical diagonalization methods \cite{Emig-exact}, and the method of numerical calculation of the Green function \cite{Johnson}. These studies have significantly advanced our understanding of the subtle effect of geometry on Casimir-Lifshitz interactions, and have led to proposals for using the knowledge in designing useful nano-scale mechanical devices \cite{machine}.

\begin{figure}[h]
\includegraphics[width=0.9\columnwidth]{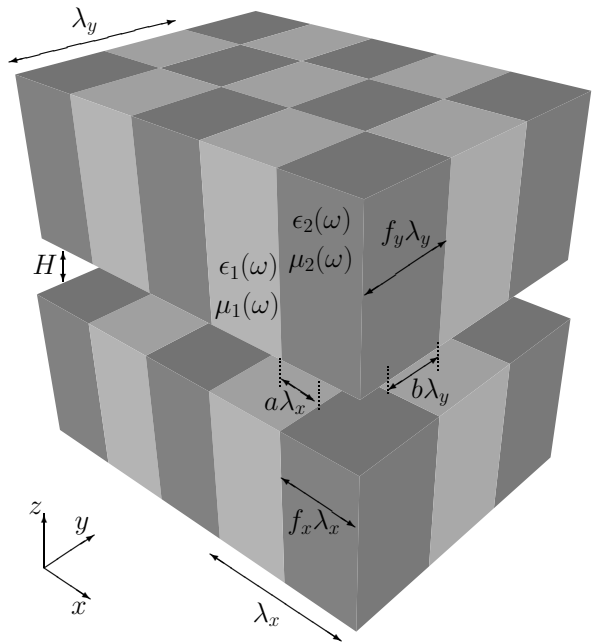}
\caption{Two semi-infinite metamaterials modeled as chessboard-patterned magneto-dielectric
media. The structure can be characterized with the separation $H$,
wavelengths $\lambda_{x}$ and $\lambda_{y}$, fractions $f_{x}$ and $f_{y}$,
permittivities $\epsilon_{1}\left(\omega\right)$ and $\epsilon_{2}\left(\omega\right)$, and
permeabilities $\mu_{1}\left(\omega\right)$ and $\mu_{2}\left(\omega\right)$. The vector $(a \lambda_{x}, b\lambda_{y})$ denotes the displacement of the upper object relative to the lower one.
} \label{fig:schem}
\end{figure}

The dependence on material properties has also been studied extensively since the work of
Dzyaloshinskii, Lifshitz, and Pitaevskii, who pointed out that the force can be attractive
or repulsive depending on the relative values of the dielectric constants of the successive
layers \cite{DLP}. The existence of a repulsive mode of the interaction is very interesting,
as it explains, for example, why a wetting layer of liquid should form on a solid in equilibrium
with vapor \cite{DLP}. Despite the theoretical possibility, it is not trivial to find a condition
where the Casimir-Lifshitz interaction between two solid bodies that are either metallic or
dielectric turns repulsive due to the presence of a non-solid medium between them. However,
a recent experiment has shown that such a repulsive force can be observed between gold and
silica particles that are separated by bromobenzene \cite{Capasso-re}. Another possibility
for repulsive Casimir-Lifshitz interactions was pointed out by Boyer, who showed that the
force between a purely dielectric semi-infinite body and a purely magnetic one separated by vacuum,
is repulsive \cite{Boyer}. Considering that the Casimir-Lifshitz force will be a key
player in the realm of micro/nano-electromechanical systems (the domain that involves length
scales of the order of 100 nm to 1 $\mu$m) the possibility of producing repulsive forces gives
hope for eliminating stiction. However, from the work of Lifshitz we know that at those distances
the Casimir-Lifshitz force will be determined by the relatively high frequency part of the
permittivity and permeability of the materials in imaginary frequency, and at such high
frequencies the response of natural magnetic materials to the electromagnetic field is
negligible ($\mu \thicksim 1$) \cite{Lifshitz,DLP,Lifshitzbook}. In other words, while the
theoretical possibility for creating a repulsive force exists, natural materials with the
required magnetic properties cannot be found.

In recent years, engineered materials---called {\em metamaterials}---have been developed based
on the proposed concept of negative refractive index \cite{veslago}, and their physical properties
have been extensively studied \cite{pendry,meta1,shalaev,soukoulis}. This development has brought about
the possibility of designing materials with special permittivity and permeability in a desirable range
of frequencies. This could, in turn, result in producing nontrivial magnetic response in a broad range of
frequencies, and possibly help achieve the repulsive Casimir-Lifshitz force \cite{kenneth,Henkel,milonni-rep,Lambrecht-pre,Soukoulis,yannop}.
A main characteristic of metamaterials is their periodic engineered structure, which could involve
features at length scales ranging from hundreds of nanometers to a few microns \cite{shalaev}.
These features could correspond to metallic split-ring resonators that are implanted in a dielectric background, or similar structures present in photonic crystals. This means that while the macroscopic response of metamaterials to electromagnetic fields can be described via appropriate frequency dependent permittivity and permeability functions \cite{pendry}, at shorter distances they should be treated as a periodic distribution of regions with contrasted permittivity and permeability response functions.
This is of particular importance in the calculation of Casimir-Lifshitz force, as we know that any lateral feature in the material properties will affect the force when the gap size is of the order of the characteristic length scale set by the heterogeneity; as argued above the typical length scales for
these features coincide with the range at which Casimir-Lifshitz forces are most significant.

The Casimir-Lifshitz force between metamaterials has been investigated recently
\cite{kenneth,Henkel,milonni-rep,Lambrecht-pre,Soukoulis,yannop}. In these studies, the
material properties are taken into consideration at the macroscopic level, in the sense that the permittivity and permeability corresponding to uniform materials have been incorporated in Lifshitz
theory. In this paper, we examine the effect of the periodic structure of metamaterials on the Casimir-Lifshitz interaction using a generalization of the dielectric contrast perturbation theory \cite{ramin,rg-09} and its application to dielectric heterostructures \cite{rah}. We develop a
perturbative scheme for the calculation of the Casimir-Lifshitz force as a series expansion in
powers of the contrast in permittivity and permeability profiles. We use the theoretical formulation to calculate the force for a model of metamaterials that is made of a two dimensional periodic structure
of varying magneto-dielectric properties, as shown in Fig. \ref{fig:schem}. We find that
the periodicity in the structure of our model metamaterials causes the normal Casimir-Lifshitz force
to change as compared to its value when the materials are assumed to be uniform. This change
is found to be significant when the distance between the two bodies is comparable to the wavelength of
periodic structure of the bodies. The heterogeneity also introduces a lateral component to the Casimir-Lifshitz
force, which is analogous to the lateral Casimir force between corrugated surfaces \cite{GK,Mohideen} and
dielectric heterostructures \cite{rah}, and is more significant at smaller separations.

The rest of the paper is organized as follows. In Sec. \ref{sec:form}, we develop the theoretical formulation of the perturbation theory that can be used for studying Casimir-Lifshitz interaction between magneto-dielectric heterostructures. Section \ref{sec:dimag} is devoted to applying the perturbative scheme to the particular problem of semi-infinite patterned magneto-dielectric structures at the leading order of the perturbation theory. In Sec. \ref{sec:norm-lat}, the results of the calculation of the normal and lateral Casimir-Lifshitz forces are shown, and finally, Sec. \ref{sec:disc} concludes the paper with some discussion and remarks.

\section{Theoretical Formulation}\label{sec:form}

\begin{figure}[t]
\includegraphics[width=0.85\columnwidth]{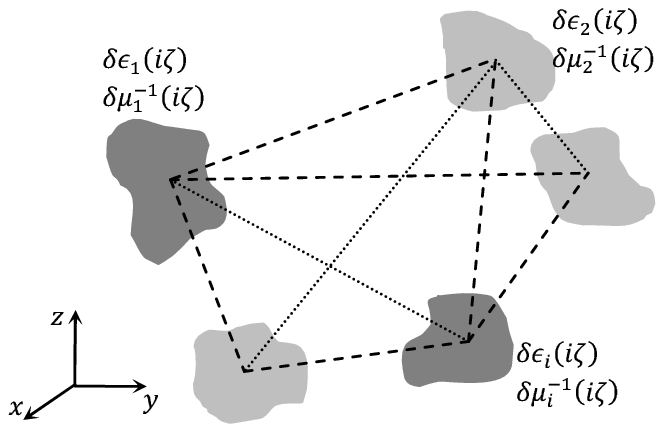}
\caption{Schematics of the assortment of magneto-dielectric objects described
by their different dielectric function and magnetic permeability profiles.} \label{fig:object}
\end{figure}

We consider an arrangement of magneto-dielectric objects in space with arbitrary shapes and frequency dependent dielectric and magnetic properties, as shown in Fig. \ref{fig:object}.
The frequency- and space-dependent dielectric function $\epsilon(\omega,{\bf r})$ and magnetic permeability $\mu(\omega,{\bf r})$ describe the medium. As shown in Ref. \cite{rg-09}, the Casimir-Lifshitz energy of the system can be written as
\begin{equation}
E_{\text{CL}}=\hbar \int_0^\infty \frac{d \zeta}{2 \pi}\; {\rm tr} \ln
\left[{\cal K}_{ij}(\zeta;{\bf r},{\bf r}')\right],\label{E-1}
\end{equation}
where
\begin{eqnarray}
{\cal K}_{ij}&=&\left[\frac{\zeta^2}{c^2} \epsilon(i \zeta,{\bf
r})\delta_{ij}+\partial_j\frac{1}{\mu\left(i \zeta,{\bf
r}\right)}\partial_i -\partial_k \frac{1}{\mu\left(i \zeta,{\bf
r}\right)}
\partial_k \delta_{ij}\right]\nonumber \\
&& \times \delta^3({\bf r}-{\bf r}'),
\label{Kij-1}
\end{eqnarray}
involves the dielectric function and magnetic permeability profiles in imaginary frequency.

%%%%%%%%%%%%%%%%%%%%%%%%%%%%%%%%%%%%%%%%%%%%%%%%%%%
\subsection{Perturbation Theory}\label{sec:pert}
%%%%%%%%%%%%%%%%%%%%%%%%%%%%%%%%%%%%%%%%%%%%%%%%%%%

Following \cite{ramin,rg-09}, we develop a systematic expansion of $E_{\text{CL}}$ in terms of the dielectric contrast
$\delta \epsilon(i \zeta,{\bf r})= \epsilon(i \zeta,{\bf r})-1 $ and the
inverse magnetic permeability contrast
$ \delta \mu ^{-1} \left(i \zeta,{\bf r}\right) = \mu^{-1} \left(i \zeta,{\bf r}\right)-1 $.
Using the Fourier transforms
\begin{eqnarray}
{\cal K}_{ij}(\zeta;{\bf q},{\bf q}') &=& \int d^{3}{\bf r} d^{3}{\bf r}' \, {\cal K}_{ij}(\zeta;{\bf r},{\bf r}')  \, e^{i {\bf q} \cdot {\bf r}}
 e^{i {\bf q}' \cdot {\bf r}'} ,\label{eq:Kqq-def} \\
\delta\tilde{\epsilon}(i \zeta,{\bf q}) &=&\int d^{3}{\bf r} \,\left[\epsilon \left(i \zeta,{\bf
r}\right)- 1\right]\, e^{i {\bf q} \cdot {\bf r}} ,\label{eq:eq-def} \\
\delta\tilde{\mu}^{-1}(i \zeta,{\bf q}) &=&\int d^{3}{\bf r} \,\left[\mu^{-1} \left(i \zeta,{\bf
r}\right)- 1\right]\, e^{i {\bf q} \cdot {\bf r}},\label{eq:muq-def}
 \end{eqnarray}
the kernel in Eq. (\ref{Kij-1}) can be decomposed as
\begin{eqnarray}
{\cal K}_{ij}(\zeta;{\bf q},{\bf q}')&=&{\cal K}_{0,ij}(\zeta,{\bf q})
(2 \pi)^3 \delta^3({\bf q}+{\bf q}') \nonumber \\
&+&\delta {\cal K}^{e}_{ij}(\zeta; {\bf q},{\bf q}')+\delta {\cal K}^{m}_{ij}(\zeta;{\bf q},{\bf q}').
\end{eqnarray}
Here
\begin{equation}
{\cal K}_{0,ij}(\zeta,{\bf q})=\frac{\zeta^2}{c^2} \delta_{ij}+q^2
\delta_{ij}-q_i q_j,
\end{equation}
corresponds to the empty space, and
\begin{eqnarray}
\delta {\cal K}^{e}_{ij}(\zeta;{\bf q},{\bf q}')&=&\frac{\zeta^2}{c^2}
\delta_{ij} \delta \tilde{\epsilon}(i \zeta,{\bf q}+{\bf
q}'),\label{dke-eq} \\
\delta {\cal K}^{m}_{ij}(\zeta;{\bf q},{\bf q}')&=&
\left(q_{j}q'_{i}-q_k q'_{k}\delta_{ij}\right)\delta
\tilde{\mu}^{-1}(i \zeta,{\bf q}+{\bf q}'),\label{dkm-eq}
\end{eqnarray}
entail the permittivity and permeability profile.

We can now recast the expression of $E_{\text{CL}}$ into a perturbative series using the identity
\begin{eqnarray}
{\rm tr} \ln [{\cal K}]&=& {\rm tr} \ln [{\cal K}_0]+\sum_{n=1}^{\infty} \frac{(-1)^{n-1}}{n}
\nonumber \\
&& \times
\; {\rm tr}\left[\left({\cal
K}_0^{-1}\,\delta {\cal K}^{e}+{\cal
K}_0^{-1}\,\delta {\cal K}^{m}\right)^{n} \right],\label{tr-eq}
\end{eqnarray}
where the inverse of the kernel ${\cal K}_0$ is given as
\begin{equation}
{\cal K}_{0,ij}^{-1}(\zeta,{\bf q})=\frac{\frac{\zeta^2}{c^2}
\delta_{ij}+q_i q_j}{\frac{\zeta^2}{c^2}
\left[\frac{\zeta^2}{c^2}+q^2\right]}.
\end{equation}

The general term for the series expansion in Eq. (\ref{tr-eq}) takes on the form
\begin{widetext}
\begin{eqnarray}
&&{\rm tr}\left[\left({\cal
K}_0^{-1}\,\delta {\cal K}^{e}+{\cal
K}_0^{-1}\,\delta {\cal K}^{m}\right)^{n} \right]=\int \frac{d^3 {\bf
q}^{(1)}}{(2 \pi)^3} \cdots \frac{d^3 {\bf q}^{(n)}}{(2 \pi)^3}\nonumber \\
&&\times \; \left[\left(\frac{\frac{\zeta^2}{c^2} \delta_{i_{1}i_{2}}+q_{i_{1}}^{(1)}
q_{i_{2}}^{(1)}}{\frac{\zeta^2}{c^2}+q^{(1)2}}\right)\delta \tilde{\epsilon}(i \zeta,-{\bf q}^{(1)}+{\bf q}^{(2)})+\left(\frac{{\bf q}^{(1)} \cdot {\bf q}^{(2)} \delta_{i_{1} i_{2}}-q^{(1)}_{i_{2}}q^{(2)}_{i_{1}}}{\frac{\zeta^2}{c^2} +q^{(1)2}}\right) \delta \tilde{\mu}^{-1}(i
\zeta,-{\bf q}^{(1)}+{\bf q}^{(2)})\right]\nonumber \\
&&\times \; \cdots \nonumber \\
&&\times \; \left[\left(\frac{\frac{\zeta^2}{c^2} \delta_{i_{n}i_{1}}+q_{i_{n}}^{(n)}
q_{i_{1}}^{(n)}}{\frac{\zeta^2}{c^2}+q^{(n)2}}\right)\delta \tilde{\epsilon}(i \zeta,-{\bf q}^{(n)}+{\bf q}^{(1)})+\left(\frac{{\bf q}^{(n)} \cdot {\bf q}^{(1)} \delta_{i_{n} i_{1}}-q^{(n)}_{i_{1}}q^{(1)}_{i_{n}}}{\frac{\zeta^2}{c^2} +q^{(n)2}}\right) \delta \tilde{\mu}^{-1}(i
\zeta,-{\bf q}^{(n)}+{\bf q}^{(1)})\right].\label{general}
\end{eqnarray}
%\end{widetext}
Using the above explicit form, the Casimir-Lifshitz energy can be calculated
for an arbitrary assortment of magneto-dielectric materials by following
standard diagrammatic methods.

In the rest of this paper, we focus only on the second order and calculate the energy for
periodic structures.

%%%%%%%%%%%%%%%%%%%%%%%%%%%%%%%%%%%%%%%%%%%%%%%%%%%
\subsection{Second Order Term}\label{sec:2nd}
%%%%%%%%%%%%%%%%%%%%%%%%%%%%%%%%%%%%%%%%%%%%%%%%%%%

We consider the leading contribution in the perturbation theory, which comes from the second-order term of the series (or its correction by a so-called Clausius-Mossotti factor using a resummation \cite{ramin,rg-09}). We note that at the second order, we can use $\delta\tilde{\mu}^{-1}=-\delta\tilde{\mu}$, which will simplify the calculations that will follow later on. We find the second order Casimir-Lifshitz energy as
%\begin{widetext}
\begin{eqnarray}
E_{2}&=&-\hbar  \int_0^\infty \frac{d \zeta}{4 \pi}
 \int \frac{d^3 {\bf k}}{(2 \pi)^3}  \frac{d^3 {\bf q}}{(2 \pi)^3} \frac{1}{(\frac{\zeta^2}{c^2}+k^{2})(\frac{\zeta^2}{c^2}+q^{2})}
 \left\{\left(3\, \frac{\zeta^4}{c^4}+ \frac{\zeta^2}{c^2} \left(k^{2}+q^{2}\right)
 +\left({\bf k} \cdot {\bf q}\right)^{2}\right)
 \,\delta \tilde{\epsilon}(i \zeta,{\bf k}+{\bf q}) \,\delta \tilde{\epsilon}(i\zeta,-{\bf k}-{\bf q})  \right. \nonumber \\
 & &  \left. +4\, \frac{\zeta^2}{c^2} \left({\bf k} \cdot {\bf q}\right)
\, \delta \tilde{\epsilon}(i \zeta,{\bf k}+{\bf q}) \,\delta\tilde{\mu}(i\zeta,-{\bf k}-{\bf q})
  +\left(\left({\bf k} \cdot {\bf q}\right)^{2}+k^{2}q^{2}\right)
  \,\delta\tilde{\mu}(i\zeta,{\bf k}+{\bf q}) \, \delta\tilde{\mu}(i\zeta,-{\bf k}-{\bf q})
  \right\}.  \label{2ndord}
 \end{eqnarray}
\end{widetext}
The above result strongly depends on the relative positioning, geometry, and electromagnetic characteristics of the bodies. Note that the $\delta\tilde{\epsilon} \delta\tilde{\mu}$ term
manifests the nonadditive nature of electric and magnetic contributions to the Casimir-Lifshitz energy.

%%%%%%%%%%%%%%%%%%%%%%%%%
\section{Magneto-dielectric Heterostructures} \label{sec:dimag}
%%%%%%%%%%%%%%%%%%%%%%%%%

We now consider two parallel semi-infinite magneto-dielectric bodies placed at a separation $H$, such as the one shown in Fig. \ref{fig:schem}. Introducing the labels {\em u} and {\em d} for ``up'' and ``down'' bodies, the permittivity function $\epsilon$ and the permeability function $\mu$ can be written as
\begin{eqnarray}
\epsilon(i \zeta,{\bf r})&=&\left\{\begin{array}{ll}
\epsilon_u(i \zeta,{\bf x}), & \;\;\;\; \frac{H}{2} \leq z < +\infty,   \\\\
1, & \;\;\;\; -\frac{H}{2} < z < \frac{H}{2},   \\\\
\epsilon_d(i \zeta,{\bf x}), & \;\;\;\; -\infty < z \leq
-\frac{H}{2},
\end{array} \right.  \\
\nonumber \\
\mu(i \zeta,{\bf r})&=&\left\{\begin{array}{ll}
\mu_u(i \zeta,{\bf x}), & \;\;\;\; \frac{H}{2} \leq z < +\infty,   \\\\
1, & \;\;\;\; -\frac{H}{2} < z < \frac{H}{2},   \\\\
\mu_d(i \zeta,{\bf x}), & \;\;\;\; -\infty < z \leq
-\frac{H}{2},
\end{array} \right.
\label{epsilon-profile}
\end{eqnarray}
where ${\bf r}= ({\bf x},z)$.

Using the identity
$$
({\zeta^2}/{c^2}+k^{2})^{-1}= \int_0^\infty dt_1 \; \exp[-t_1 ({\zeta^2}/{c^2}+k^{2})],
$$
%and $ ({\zeta^2}/{c^2}+q^{2})^{-1}= \int_0^\infty dt_2 \exp[-t_2 ({\zeta^2}/{c^2}+q^{2})]$,
and introducing the new variables ${\bf P} \equiv \frac{1}{2}({\bf k}-{\bf q})$ and
${\bf Q} \equiv {\bf k}+{\bf q}$, we can simplify Eq. (\ref{2ndord}) by performing the
Gaussian integrations over the variable ${\bf P}$. This yields
\begin{eqnarray}
E_{2}&=&-\frac{\hbar}{4 \pi^2 c^2} \int_0^\infty d \zeta \; \zeta^2 \int
d^2{\bf x} d^2{\bf x}' \int \frac{d^2 {\bf Q}_{\bot}}{(2 \pi)^2}
\;{\rm e}^{i {\bf Q}_{\bot} \cdot ({\bf x}-{\bf x}')} \nonumber \\
&\times& \int_{1}^{\infty}d p \; \frac{{\rm e}^{-\frac{\zeta
H}{c}\sqrt{4 p^{2}+\left( c
{Q}_{\bot}/\zeta\right)^{2}}}}{\left[4 p^{2}+\left( c
{Q}_{\bot}/\zeta\right)^{2}\right]^{3/2}}\;\left[   \mathcal{E}_{ee}+    \mathcal{E}_{em}  +  \mathcal{E}_{mm} \right], \nonumber
\\\label{E2}
\end{eqnarray}
where
\begin{eqnarray}
\mathcal{E}_{ee}&=& (2p^{4}-2p^{2}+1)  \delta \epsilon_u(i\zeta,{\bf x}) \delta \epsilon_d(i \zeta,{\bf x}'),\label{Eee-def}  \\
\mathcal{E}_{em}&=&(-2p^{2}+1)  \label{Eem-def} \\
&\times & \Bigl(\delta \epsilon_u(i\zeta,{\bf x}) \delta \mu_d(i \zeta,{\bf x}')+
\delta \mu_u(i \zeta,{\bf x}) \delta\epsilon_d(i \zeta,{\bf x}')  \Bigr),\nonumber \\
\mathcal{E}_{mm}&=& (2p^{4}-2p^{2}+1)  \delta \mu_u(i\zeta,{\bf x}) \delta \mu_d(i \zeta,{\bf x}'). \label{Emm-def}
\end{eqnarray}
This result can now be used to study the Casimir-Lifshitz interaction between two macroscopic
bodies with any magneto-dielectric profiles at the second order in perturbation theory.

%%%%%%%%%%%%%%%%%%%%%%%%%%%%%%%%%%%%%%%%%%%%%%%%%%%
\subsection{Two homogenous semi-infinite bodies}\label{sec:dimag1}
%%%%%%%%%%%%%%%%%%%%%%%%%%%%%%%%%%%%%%%%%%%%%%%%%%%

Let us first consider two homogenous semi-infinite bodies of area $A$ and
separation $H$. Assuming that the permittivity and permeability are frequency-independent, the
Casimir-Lifshitz force $F=-{\partial  E_2 }/{\partial H}$ takes on a simple form
\begin{eqnarray}
F&=&-\frac{\hbar c A}{640  \pi^2 H^4} \nonumber \\
&\times& \left[23\left(\delta\epsilon_d\delta
\epsilon_u+\delta \mu_d \delta \mu_u\right)-7\left(\delta
\epsilon_d \delta \mu_u+ \delta
\epsilon_u \delta \mu_d\right) \right],\nonumber \\
\end{eqnarray}
as found previously in the literature \cite{Casimir-P,Fei-Suc,Lubkin}.
%Here the nonadditive nature of "electric" and "magnetic" contributions to
%the Casimir-Lifshitz force is clear.
We can now consider patterned magneto-dielectric objects, and study how the structural
heterogeneity affects both the normal and lateral Casimir forces.

%%%%%%%%%%%%%%%%%%%%%%%%%
\subsection{Patterned magneto-dielectric objects}\label{sec:dimag2}
%%%%%%%%%%%%%%%%%%%%%%%%%

\begin{figure}[h]
\includegraphics[width=0.8\columnwidth]{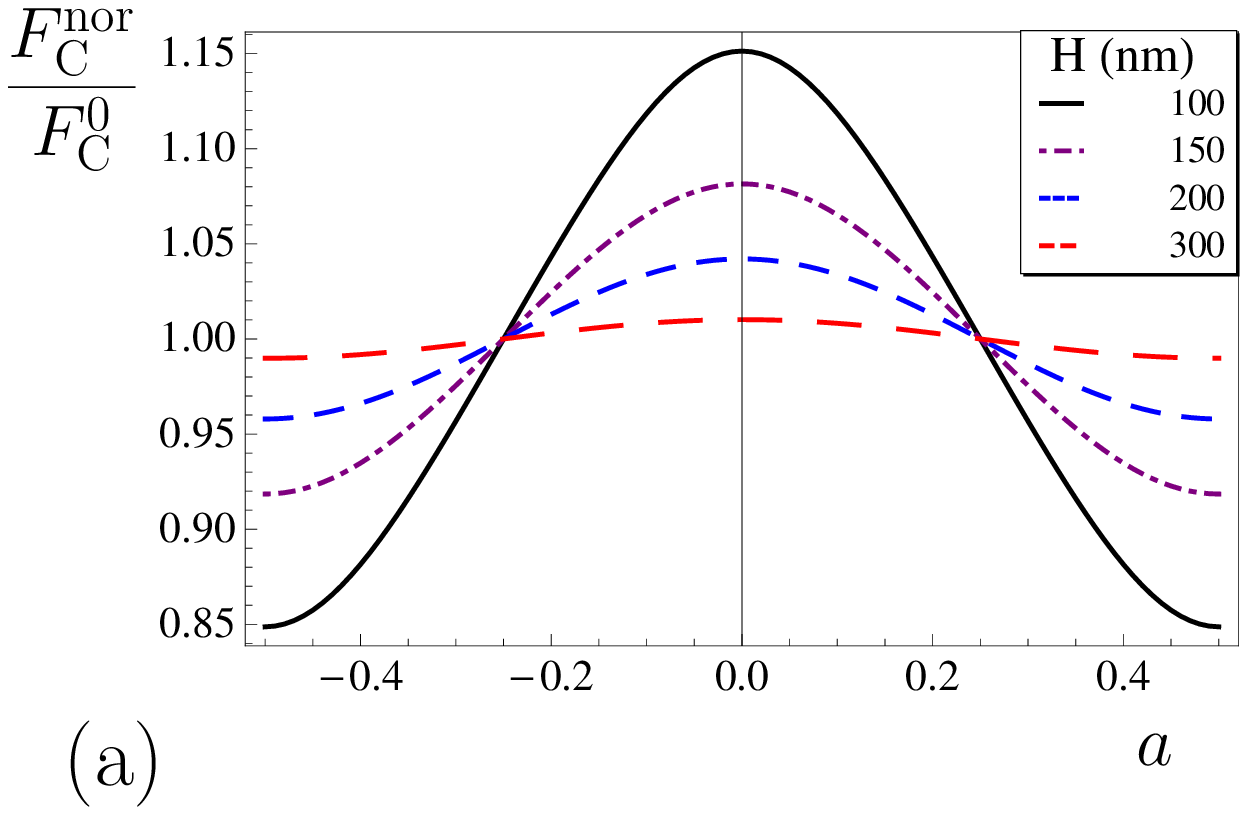}\\
\vspace{.5cm}
\includegraphics[width=0.8\columnwidth]{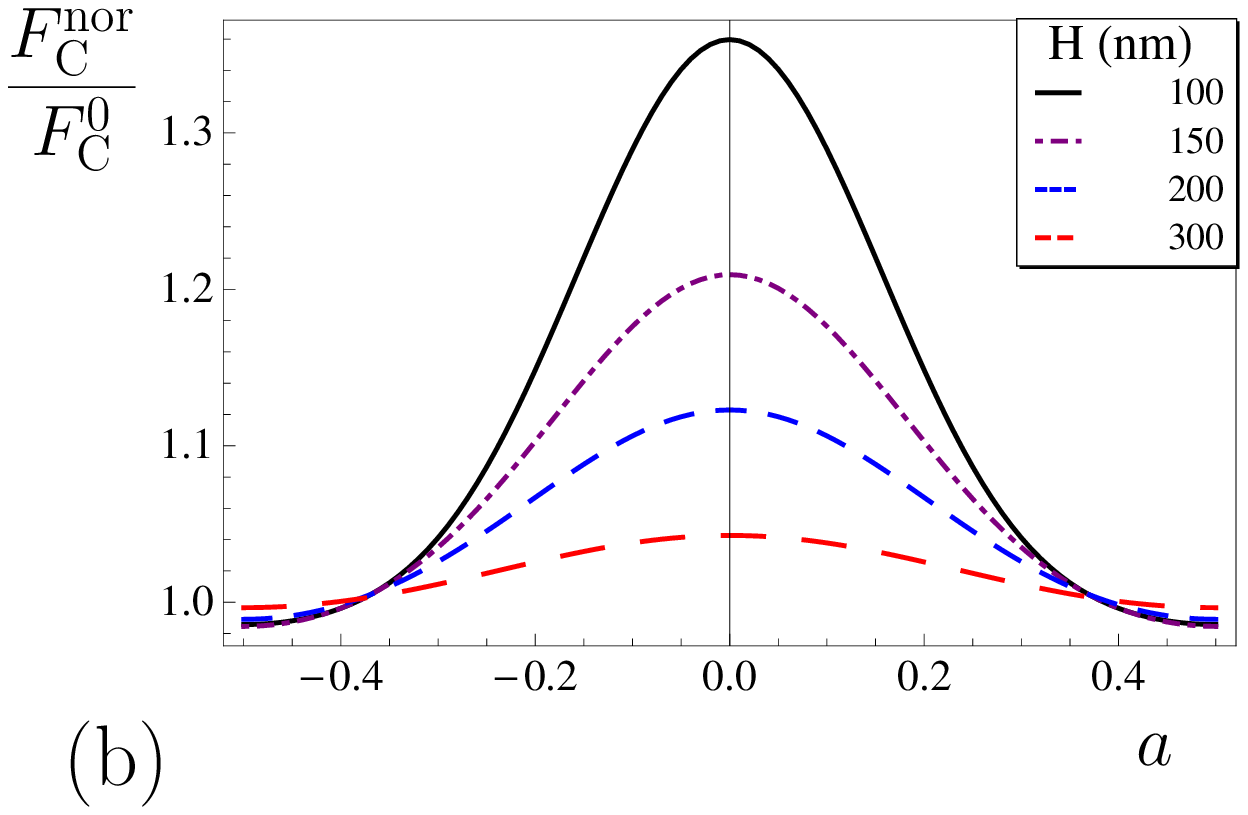}\\
\caption{(Color online). The normal Casimir force (in units of $F^{0}_{C}$) for the case of
{\em EhMh--ElMl}, as a function of the dimensionless parameter $a$ for various distances $H$.
Here $\lambda_{x}=\lambda_{y}= 500$ nm, and the two panels correspond to
(a) $f_{x}=f_{y}=0.5$, and (b) $f_{x}=0.75$ and $f_{y}=0.25$.
} \label{fig:new-normal}
\end{figure}

\begin{figure}[h]
\includegraphics[width=0.8\columnwidth]{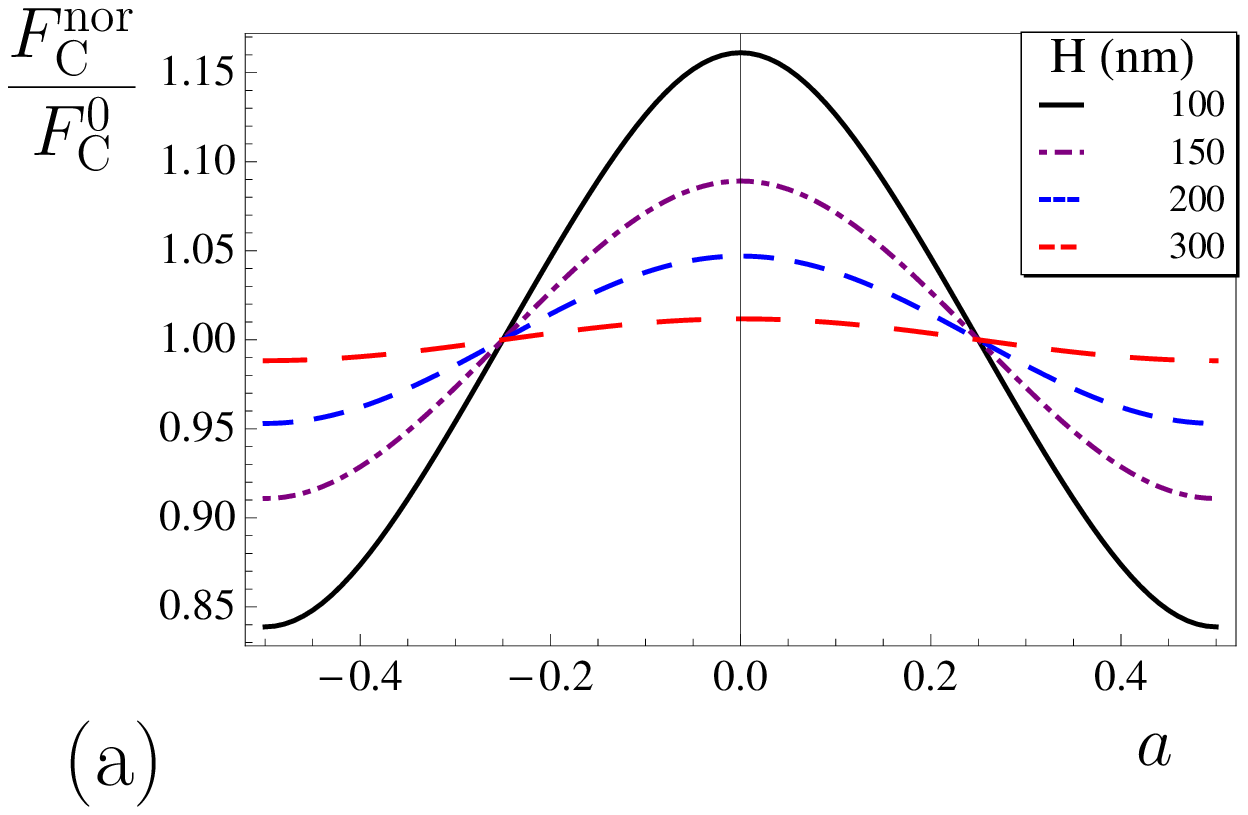}\\
\vspace{.5cm}
\includegraphics[width=0.8\columnwidth]{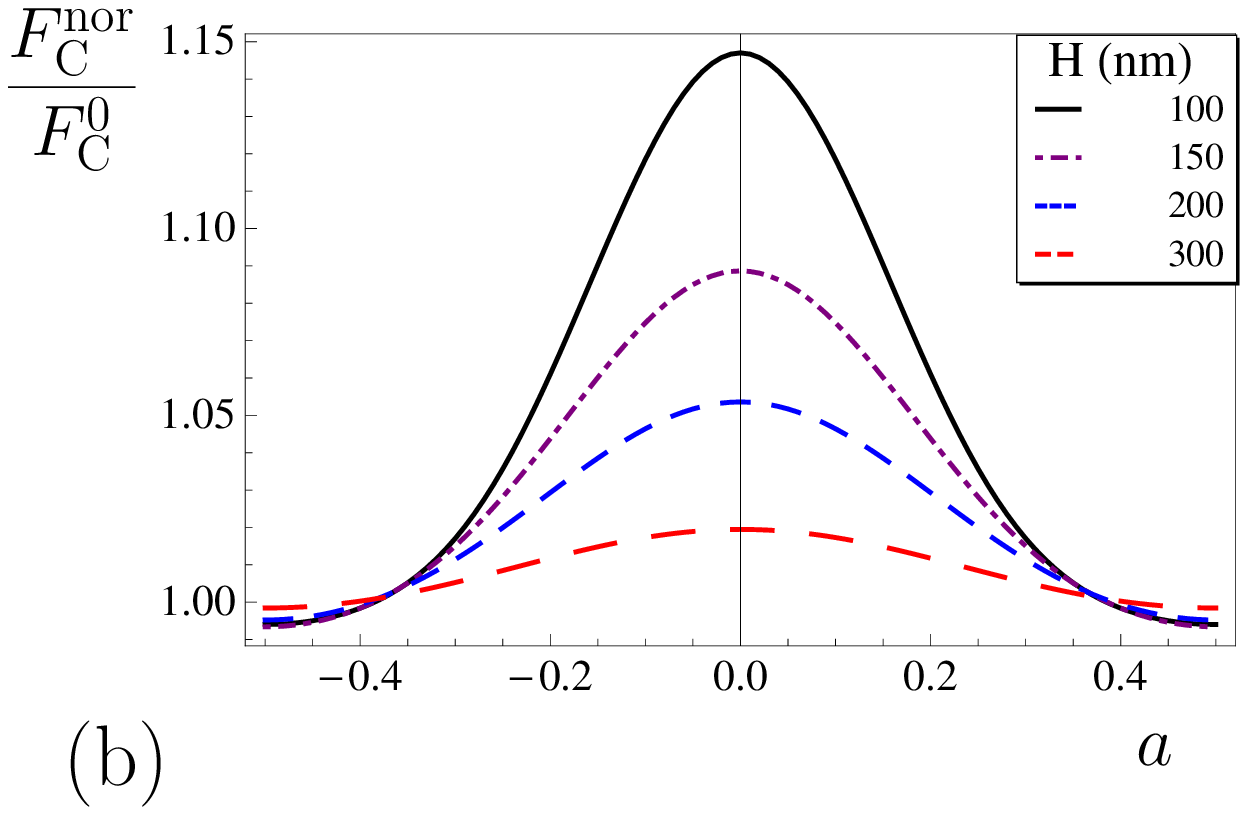}\\
\caption{(Color online). The normal Casimir force (in units of $F^{0}_{C}$) for the case of
{\em ElMh--EhMl}, as a function of the dimensionless parameter $a$ for various distances $H$.
Here $\lambda_{x}=\lambda_{y}= 500$ nm, and the two panels correspond to
(a) $f_{x}=f_{y}=0.5$, and (b) $f_{x}=0.75$ and $f_{y}=0.25$.
} \label{fig:normal}
\end{figure}

Let us now consider the configuration shown in Fig. \ref{fig:schem}.
The magneto-dielectric ``chessboard'' heterostructure can be characterized
with wavelengths $\lambda_{x}$ and $\lambda_{y}$ along the $x$ and $y$ directions.
In a repeat unit of the material along each direction $\alpha$ ($\alpha=x,y$),
a fraction $f_\alpha$ of the material has permittivity $\epsilon_{2}\left(\omega\right)$ and permeability $\mu_{2}\left(\omega\right)$, and the remaining fraction $(1-f_\alpha)$ has
permittivity $\epsilon_{1}\left(\omega\right)$ and permeability $\mu_{1}\left(\omega\right)$.
Below, we will consider two possibilities; one with the domains with higher permittivity and
permeability coinciding and another where they are in a staggered configuration.
The vector $(a \lambda_{x}, b\lambda_{y})$ denotes the displacement of the upper object
relative to the lower one, as shown in Fig. \ref{fig:schem}.

We use the Clausius-Mossotti resummation of the perturbation theory for the permittivity
contribution \cite{ramin,rg-09}, which amounts to replacing $\delta \epsilon(i \zeta,{\bf x})$
in Eq. (\ref{E2}) by
$$
\overline{\delta \epsilon}(i \zeta,{\bf x})=
\left[\frac{\delta \epsilon(i \zeta,{\bf x})}{1+\frac{1}{3} \delta
\epsilon(i \zeta,{\bf x})}\right].
$$
Due to the periodicity of the magneto-dielectric structures, it is natural to use the Fourier series expansion of the permittivity and permeability profiles. We have
\begin{eqnarray}
&&\overline{\delta \epsilon_d}(i \zeta,{\bf x})=\sum_{n,m=-\infty}^{\infty} {\mathcal A}_{nm}(i \zeta) \;{\rm e}^{i 2 \pi n \frac{x}{\lambda_{x}} +i 2 \pi m \frac{y}{\lambda_{y}}}, \nonumber \\
&&\overline{\delta \epsilon_u}(i \zeta,{\bf x})=\sum_{n,m=-\infty}^{\infty} {\mathcal A}_{nm}(i \zeta)\;{\rm e}^{i 2 \pi n \frac{(x +a \lambda_{x} )}{\lambda_{x}}+i 2 \pi m \frac{(y+  b\lambda_{y})}{\lambda_{y}}}, \nonumber \\
&&{\delta \mu_d}(i \zeta,{\bf x})=\sum_{n,m=-\infty}^{\infty} {\mathcal B}_{nm}(i \zeta)\;
{\rm e}^{i 2 \pi n \frac{x}{\lambda_{x}} +i 2 \pi m \frac{y}{\lambda_{y}}}, \nonumber \\
&&{\delta \mu_u}(i \zeta,{\bf x})=\sum_{n,m=-\infty}^{\infty} {\mathcal B}_{nm}(i \zeta)\;
{\rm e}^{i 2 \pi n \frac{(x +a \lambda_{x})}{\lambda_{x}}+i 2 \pi m \frac{(y+  b\lambda_{y})}{\lambda_{y}}}. \nonumber
\end{eqnarray}
The Fourier series coefficients can be easily found as
\begin{equation}
{\mathcal A}_{nm}(i \zeta)= 2  \left[\overline{\delta \epsilon_2}(i \zeta)-\overline{\delta \epsilon_1}(i \zeta)\right]\;\frac{\sin(n \pi f_{x}) \sin(m \pi f_{y})}{m n \pi^{2}},
\label{Adef}
\end{equation}
and
\begin{equation}
{\mathcal B}_{nm}(i \zeta)= 2  \left[{\delta \mu_2}(i \zeta)-{\delta \mu_1}(i \zeta)\right]\;
\frac{\sin(n \pi f_{x}) \sin(m \pi f_{y})}{m n \pi^{2}},
\label{Bdef}
\end{equation}
for $ n\neq 0$ and $ m\neq 0$.
Using Eq. (\ref{E2}), one obtains the Casimir-Lifshitz energy $E_{\text{C}}$ of the chessboard
magneto-dielectric heterostructure as
\begin{eqnarray}
E_{\text{C}}&=&-\frac{\hbar A}{2\pi^{2} c^2}{\sum_{m,n=0}^{\infty}}^{'}
\; \cos\left(2 \pi n a + 2 \pi m b \right) \nonumber \\
&\times &\int_{0}^{\infty} d\zeta \int_{1}^{\infty} d p \;
\zeta^2 \;\frac{{\rm e}^{-\frac{\zeta H}{c}\sqrt{4 p^{2}+\left(\frac{2 \pi n c}{\lambda_{x} \zeta}\right)^2+\left(\frac{2 \pi m c}{\lambda_{y} \zeta}\right)^2}}}{\left[4 p^{2}+\left(\frac{2 \pi n c}{\lambda_{x} \zeta}\right)^2+\left(\frac{2 \pi m c}{\lambda_{y} \zeta}\right)^2\right]^{3/2}} \nonumber \\
&& \times \left[(2p^{4}-2p^{2}+1)  \Bigl({\cal A}_{n m
}^{2}\left(i \zeta\right)+ {\cal B}_{n m}^{2}\left(i \zeta\right)
\Bigr) \right.  \nonumber \\
& & \left. -2 (2 p^{2}-1) {\cal A}_{n m}\left(i \zeta\right) {\cal B}_{n m}\left(i \zeta\right)\right],\label{Epp}
\end{eqnarray}
where the prime on the summation indicates that the $m=n=0$ term comes with a
prefactor of $1/2$.

%%%%%%%%%%%%%%%%%%%%%%%%%
\subsection{Material Properties}\label{sec:material}
%%%%%%%%%%%%%%%%%%%%%%%%%

\begin{figure}[h]
\includegraphics[width=0.8\columnwidth]{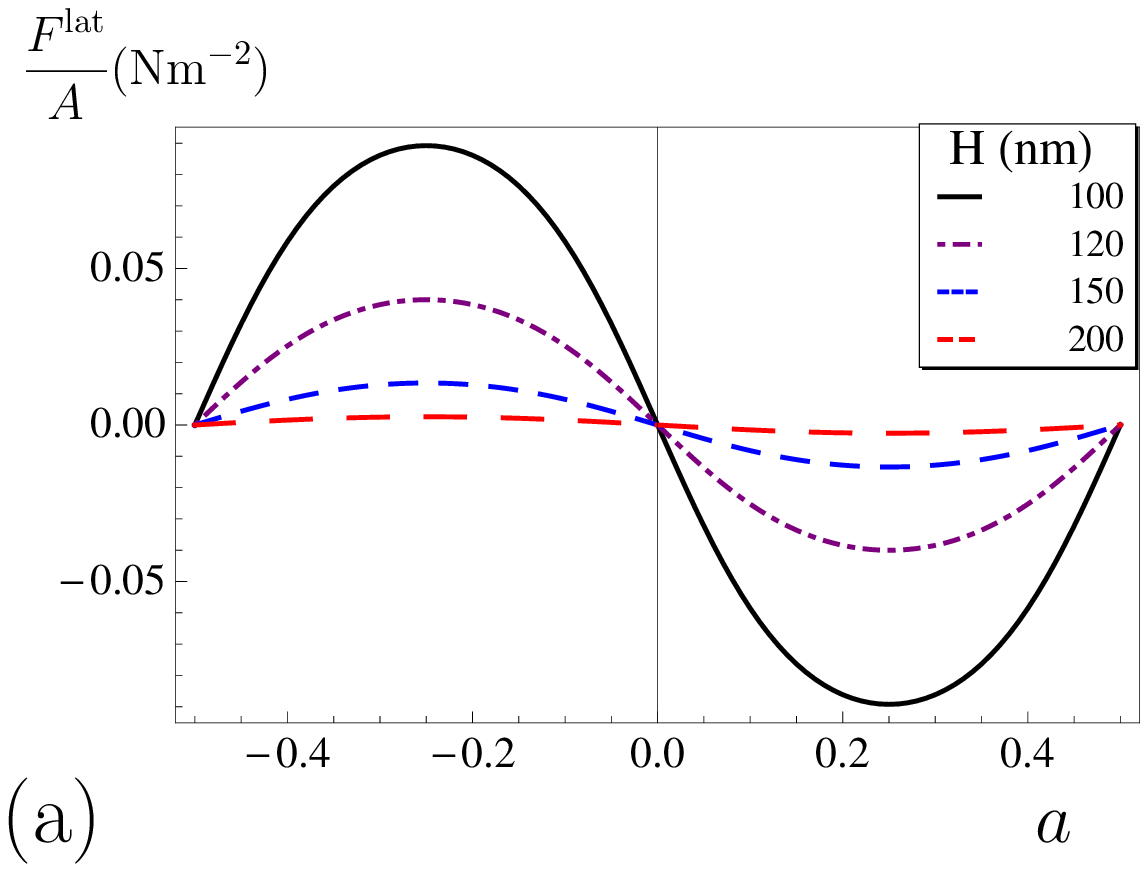}\\
\vspace{.5cm}
\includegraphics[width=0.8\columnwidth]{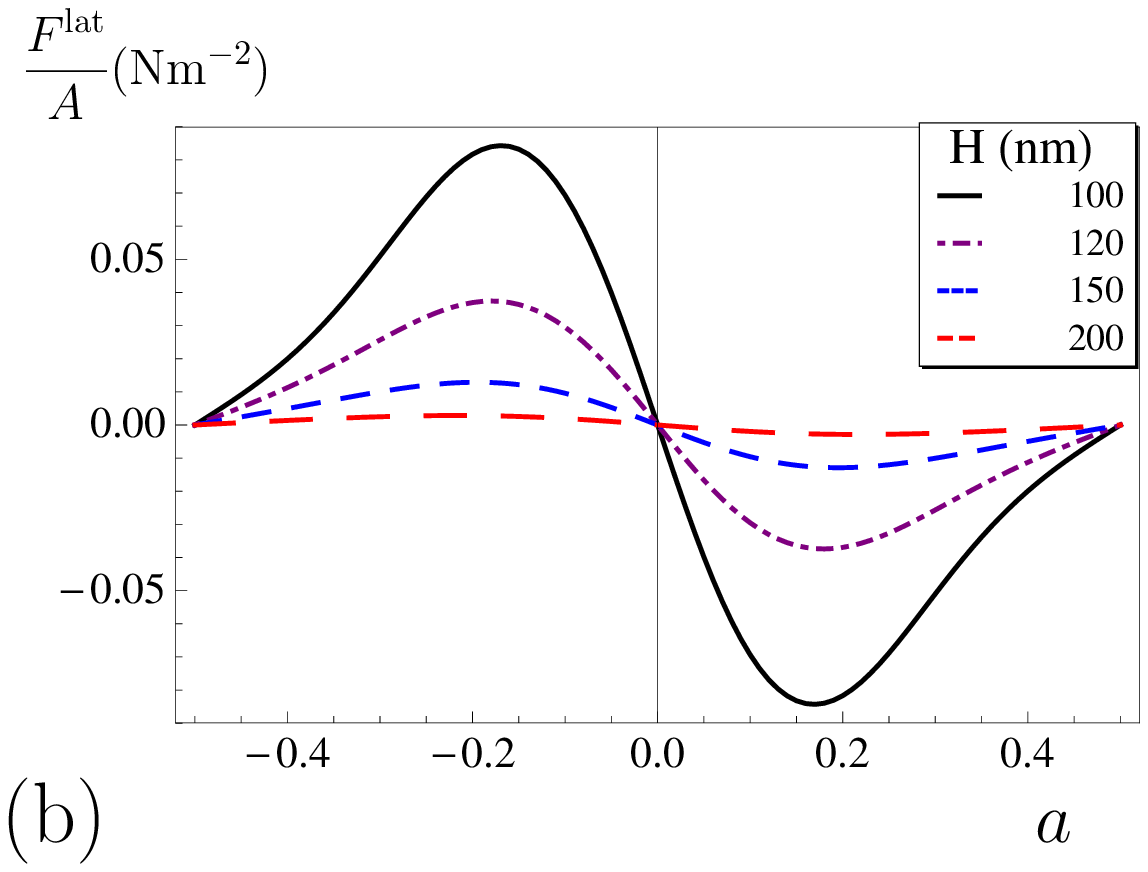}\\
\caption{(Color online). The lateral Casimir force for the case of {\em EhMh--ElMl},
as a function of the dimensionless parameter $a$ for various distances $H$.
Here $\lambda_{x}=\lambda_{y}= 500$ nm, and the two panels correspond to
(a) $f_{x}=f_{y}=0.5$, and (b) $f_{x}=0.75$ and $f_{y}=0.25$.
} \label{fig:new-lateral}
\end{figure}

\begin{figure}[h]
\includegraphics[width=0.8\columnwidth]{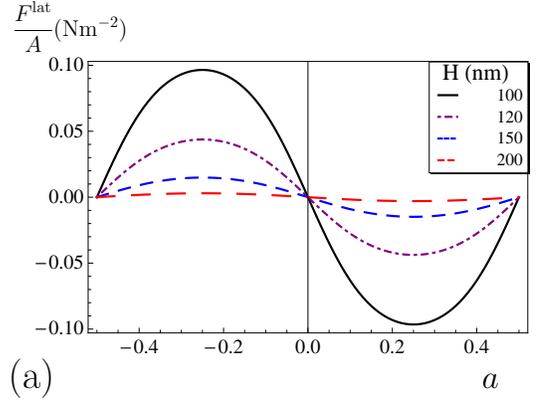}\\
\vspace{.5cm}
\includegraphics[width=0.8\columnwidth]{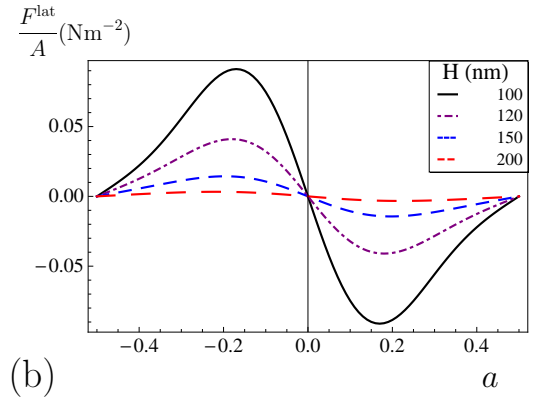}\\
\caption{(Color online). The lateral Casimir force for the case of
{\em ElMh--EhMl}, as a function of the dimensionless parameter $a$ for various distances $H$.
Here $\lambda_{x}=\lambda_{y}= 500$ nm, and the two panels correspond to
(a) $f_{x}=f_{y}=0.5$, and (b) $f_{x}=0.75$ and $f_{y}=0.25$.
} \label{fig:lateral}
\end{figure}

While the experimental realizations of metamaterials involve a multitude of complex structures,
in the present study we consider a simplified model where the heterostructure could have two types
of effective permittivities---corresponding to metallic and dielectric materials---and two types of
permeabilities---corresponding to magnetic and non-magnetic materials.
For the metal, which has a relatively {\em higher} dielectric constant especially at lower frequencies,
we use the Drude model that in imaginary frequency reads
\begin{equation}
\epsilon_h(i \zeta)=1+\frac{\Omega_D^2}{\zeta^2 + \gamma_D \zeta}. \label{eq:eh}
\end{equation}
Alternatively, for the dielectric medium with relatively {\em lower} dielectric function we use the
Drude-Lorentz model
\begin{equation}
\epsilon_l(i \zeta)=1+\frac{\Omega_{e}^2}{\zeta^2+\omega_{e}^2+\gamma_{e} \zeta}.\label{eq:el}
\end{equation}
Similarly, we choose a simple Drude-Lorentz model for the permeability of the magnetic material,
namely
\begin{equation}
\mu_h(i \zeta)=1+\frac{\Omega_{m}^2}{\zeta^2+\omega_{m}^2+\gamma_{m}\zeta},\label{eq:mh}
\end{equation}
while for the non-magnetic material we have
\begin{equation}
\mu_l(i \zeta )=1.\label{eq:mh}
\end{equation}
In the above equations, $\omega_{e}$ ($\omega_{m}$) is the electric (magnetic) resonance frequency,
and $\gamma_{e}$ ($\gamma_{m}$) is the electric (magnetic) dissipation parameter.% \cite{pendry}.
Using the plasma frequency of gold $\omega_{p}({\rm Au})\equiv\omega_{p}=1.37 \times 10^{16}$ rad/s as a
frequency scale, the numerical values of the magneto-dielectric characteristic parameters
are chosen as:
$\displaystyle{\Omega_{D}/\omega_{p}=1.0}$,
$\displaystyle{\gamma_{D}/\omega_{p}=0.004}$,
$\displaystyle{\Omega_{e}/\omega_{p}=0.04}$,
$\displaystyle{\Omega_{m}/\omega_{p}=0.1}$,
$\displaystyle{\omega_{e}/\omega_{p}=\omega_{m}/\omega_{p}=0.1}$, and
$\displaystyle{\gamma_{e}/\omega_{p}=\gamma_{m}/\omega_{p}=0.005}$ \cite{milonni-rep}.

We consider two different possibilities: (1) When the metallic patch has magnetic properties and the dielectric patch is non-magnetic. In this case, which we represent it schematically as {\em EhMh--ElMl}, we have $\epsilon_2=\epsilon_h$, $\mu_2=\mu_h$, $\epsilon_1=\epsilon_l$, and $\mu_1=\mu_l$. (2) When the dielectric patch has magnetic properties and the metallic patch is non-magnetic. In this case, which we represent it schematically as {\em ElMh--EhMl}, we have $\epsilon_2=\epsilon_l$, $\mu_2=\mu_h$, $\epsilon_1=\epsilon_h$, and $\mu_1=\mu_l$. Below, we will study both the normal and lateral Casimir-Lifshitz forces in both of these cases.

\begin{figure*}
\includegraphics[width=0.63\columnwidth]{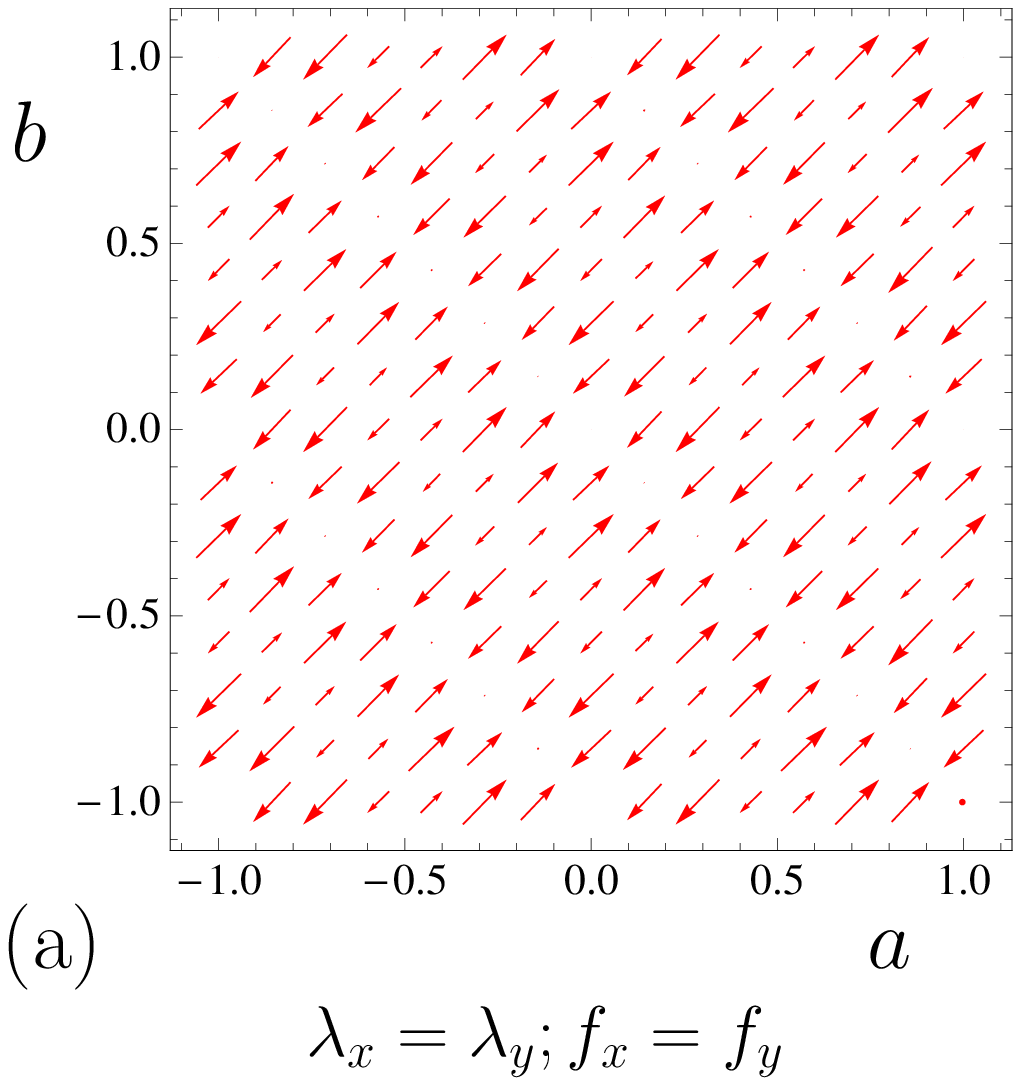}
\hspace{.5cm}
\includegraphics[width=0.63\columnwidth]{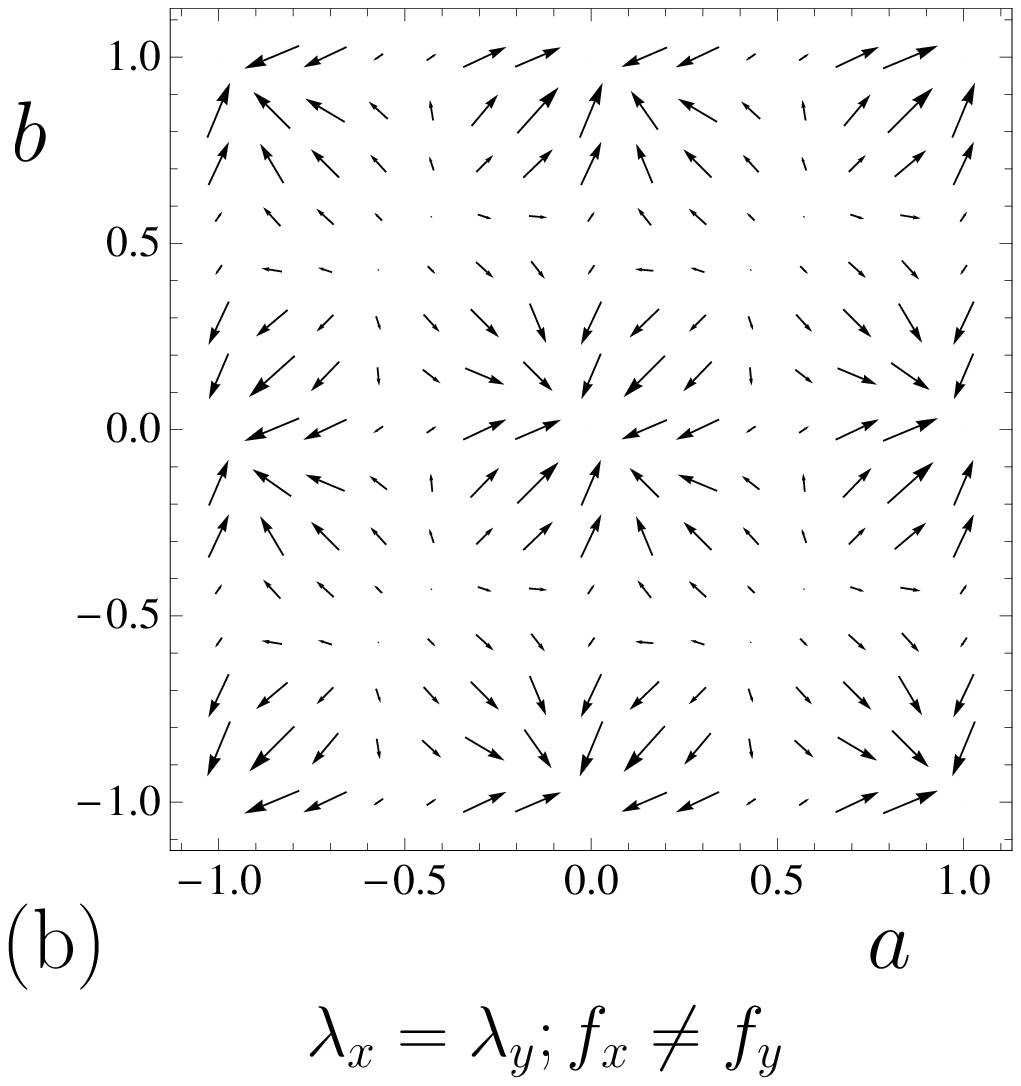}
\hspace{.5cm}
\includegraphics[width=0.63\columnwidth]{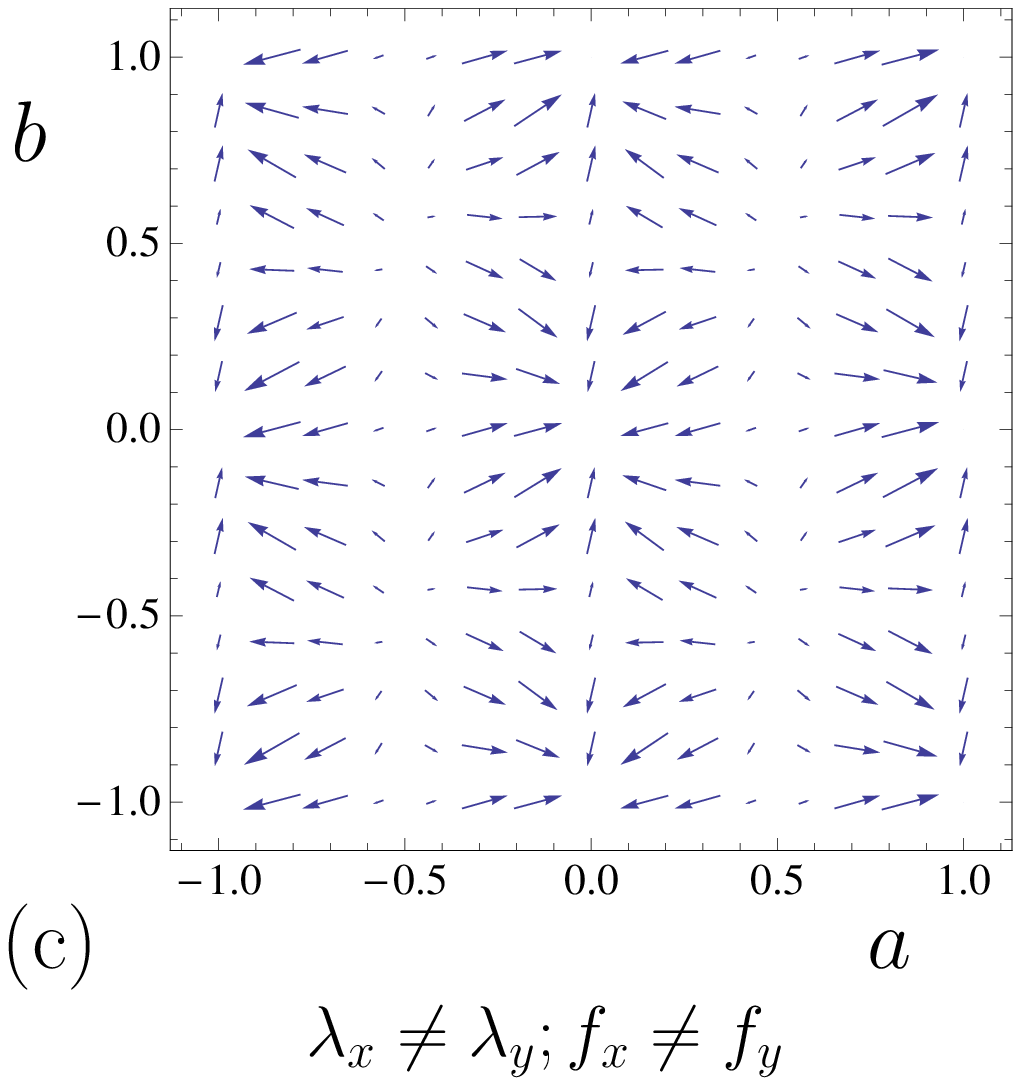}
\caption{
(Color online). Vector field plots of the lateral Casimir-Lifshitz force as a function of the relative displacements $a$ and $b$, along the $x$ and $y$ directions respectively, for the EhMh-ElMl model at $H=100$ nm. (a) $f_{x}=f_{y}=0.5$ and $\lambda_{x}=\lambda_{y}= 500$ nm. (b) $f_{x}=0.75$ and $f_{y}=0.25$ and $\lambda_{x}=\lambda_{y}= 500$ nm. (c) $f_{x}=0.75$, $f_{y}=0.25$, $\lambda_{x}=500$ nm, and $\lambda_{y}= 200$ nm.
} \label{fig:vec}
\end{figure*}

\section{Casimir-Lifshitz Forces between the Chessboard Structures} \label{sec:norm-lat}

Using the material properties described in Sec. \ref{sec:material} above, we can now calculate the Casimir-Lifshitz energy for the chessboard magneto-dielectric heterostructure shown in Fig. \ref{fig:schem}
for the two cases denoted as {\em EhMh--ElMl} and {\em ElMh--EhMl} above. Due to the lateral heterogeneity
of the magneto-dielectric properties, the two bodies exert both normal and lateral Casimir-Lifshitz forces
on each other.

\subsection{Normal Force} \label{sec:norm}

The normal force between the two patterned structures can be calculated as
$F^{\text{nor}}_{\text{C}}=-\frac{\partial E_{\text{C}}}{\partial H}$. We set
$b=0$, and study the normal force as a function of $a$ for various values of
the gap size $H$. We choose to normalize the force using $F^{0}_{C}$, which is the
contribution of the $m=n=0$ term in Eq. (\ref{Epp}) to the normal force.
Figure \ref{fig:new-normal} shows the normal force relative to $F^{0}_{C}$
for the {\em EhMh--ElMl} case. Figure \ref{fig:new-normal}a corresponds to
the symmetric configuration where $f_{x}=f_{y}=0.5$ whereas Fig. \ref{fig:new-normal}b
corresponds to an asymmetric configuration where $f_{x}=0.75$ and $f_{y}=0.25$, and
in both cases $\lambda_{x}=\lambda_{y}= 500$ nm. Figure \ref{fig:new-normal} shows
that depending on how the different patches with different magneto-dielectric properties
are positioned with respect to one another, the normal Casimir-Lifshitz force can
change in magnitude relative to the case with uniform magneto-dielectric configuration.
When patches with similar properties are opposite one another ($a=0$) the attractive normal
force is at its maximum, while the force is weakest when dissimilar patches are exactly opposite
one another. While this relative change depends strongly on the gap size, we note that it can
easily amount to a few percent in the experimentally relevant gaps sizes of a few hundred
nanometers, and could even reach the value of 35\% for the gap size of $H=100$ nm
for the asymmetric example. Figure \ref{fig:normal} shows a similar behavior for
the {\em ElMh--EhMl} case, which shows a relatively less dramatic change in the
asymmetric example.

\subsection{Lateral Force} \label{sec:lat}

The lateral force between the two patterned structures is a vector, with its value and direction
depending on the relative positioning of the two bodies. For simplicity, we first focus on the case
with $b=0$, and only study the lateral force for unidirectional displacements along the $x$ axis
(see Fig. \ref{fig:schem}). In this case, the force also lies along the $x$ axis for symmetry reasons,
and we can find its value using $F^{\text{lat}}=-\frac{1}{\lambda_{x}}\frac{\partial E_{\text{C}}}{\partial a}$,
as a function of $a$ for different gap sizes $H$. Figure \ref{fig:new-lateral} shows the lateral
force per unit area in SI units, for the {\em EhMh--ElMl} case.
Figure \ref{fig:new-lateral}a corresponds to the symmetric configuration where $f_{x}=f_{y}=0.5$
whereas Fig. \ref{fig:new-lateral}b corresponds to an asymmetric configuration
where $f_{x}=0.75$ and $f_{y}=0.25$, and in both cases $\lambda_{x}=\lambda_{y}= 500$ nm.
Figure \ref{fig:new-lateral} shows that the lateral Casimir-Lifshitz force is very sensitive to the
value of $H$, and its dependence on the lateral displacement reflects the symmetry or asymmetry of the
relative sizes of the two patches. While the form of the lateral force at relatively larger gap sizes
tends to a sinusoidal form, at smaller separations higher harmonics contribute as well to reflect
more of the details of the heterogeneity. Figure \ref{fig:lateral} shows the lateral force for
the {\em ElMh--EhMl} case, which shows a similar behavior as compared to the previous case.

In Fig. \ref{fig:vec}, the vector field for the lateral Casimir-Lifshitz force, defined as ${\bf F}^{\text{lat}}(a,b)=-\frac{1}{\lambda_{x}}\frac{\partial E_{\text{C}}}{\partial a} {\hat {\bf x}}-\frac{1}{\lambda_{y}}\frac{\partial E_{\text{C}}}{\partial b} {\hat {\bf y}}$, is plotted as a function
of $a$ and $b$. One can see that the symmetry of the heterostructure affects the configuration of the lateral Casimir-Lifshitz force as a vector field, and that the numerous parameters involved can provide opportunities for a rich variety of engineered patterns for the lateral force.

\section{Discussion}\label{sec:disc}

We have studied the Casimir-Lifshitz interaction between two metamaterials modeled as
periodic arrayed structures containing domains of varying magneto-dielectric properties.
We have considered two types of permittivity functions---metallic and dielectric---and two
types of permeability functions---magnetic and non-magnetic---and their corresponding two
combinations. For both combinations, we have found significant changes in the value of the normal
Casimir-Lifshitz force relative to the value that corresponds to the uniform (macroscopic) model of the
materials. The relative change is increased as the gaps size is decreased, and reaches
a few percent in the experimentally relevant gaps sizes of a few hundred nanometers, while
it could even reach 35\% when the gap size is $100$ nm in the models we studied. Considering
how delicate it is to find the condition to achieve the repulsive force in realistic situations as
recent studies have revealed \cite{milonni-rep}, our results show that the effect of the structural
heterogeneity should be taken into account in determining whether the force is repulsive
or not. This is particularly pertinent as the characteristic length scale of the periodic features
in realistic metamaterials coincides with the gap sizes at which the Casimir-Lifshitz force is particularly
relevant. While this issue has been ignored by all previous studies, we note that our study has been
performed within the scope of the magneto-dielectric contrast perturbation theory at its leading order
and should be considered more as an indication of the relative significance of the effect rather than
a study that could provide numerically accurate results \cite{note}. To that end, one needs to employ more
sophisticated numerical methods similar to those developed to study the effect of geometry
\cite{Emig-exact,Johnson}.

Another consequence of the structural heterogeneity of the metamaterials is the possibility of the
emergence of lateral Casimir-Lifshitz forces, which are stronger for smaller gap sizes. While all previous works
on lateral Casimir force have focused only on unidirectional geometrical or material heterostructure features,
we have considered a two dimensional pattern and presented the vector field distribution of the lateral force.
These forces are very sensitive to the details of the periodic patterns of the magneto-dielectric
properties, and their versatility allows them to be amenable to detailed engineering by changing these features.

\acknowledgements

The authors wish to thank the ESF Research Network CASIMIR for providing
excellent opportunities for discussion on the Casimir effect and
related topics. This work was supported by the EPSRC under Grants
EP/E024076/1 and EP/F036167/1.


\begin{thebibliography}{99}
\bibitem{note}
For example, we do not observe a repulsive force in the regime of parameters
where Lifshitz theory does show repulsion, as found in Ref. \cite{milonni-rep}.
\end{thebibliography}
\end{document}